\documentclass[twocolumn,aps,prl,superscriptaddress,showpacs,floatfix]{revtex4}
\usepackage{graphicx}

\begin{document}

\title{Effects of symmetry energy on two-nucleon correlation 
functions in heavy-ion collisions induced by neutron-rich nuclei}
\author{Lie-Wen Chen}
\thanks{On leave from Department of Physics, Shanghai Jiao Tong University,
Shanghai 200030, China}
\affiliation{Cyclotron Institute and Physics Department, Texas A\&M 
University, College Station, Texas 77843-3366}
\author{Vincenzo Greco}
\affiliation{Cyclotron Institute and Physics Department, Texas A\&M 
University, College Station, Texas 77843-3366}
\author{Che Ming Ko}
\affiliation{Cyclotron Institute and Physics Department, Texas A\&M 
University, College Station, Texas 77843-3366}
\author{Bao-An Li}
\affiliation{Department of Chemistry and Physics, P.O. Box 419, Arkansas State
University, State University, Arkansas 72467-0419}
\date{\today}

\begin{abstract}
Using an isospin-dependent transport model, we study the effects of nuclear 
symmetry energy on two-nucleon correlation functions in heavy ion collisions
induced by neutron-rich nuclei. We find that the density dependence of the 
nuclear symmetry energy affects significantly the nucleon emission times 
in these collisions, leading to larger values of two-nucleon correlation 
functions for a symmetry energy that has a stronger density dependence. 
Two-nucleon correlation functions are thus useful tools for extracting
information about the nuclear symmetry energy from heavy ion collisions.
\end{abstract}

\maketitle

The nuclear symmetry energy is a measure of the difference between the 
binding energy of an asymmetric nuclear matter and that of a nuclear matter 
with equal numbers of protons and neutrons. Knowledge on the 
density-dependence of the nuclear symmetry energy is essential for 
understanding not only the structure of radioactive nuclei 
\cite{oya,brown,hor01,furn02} but also many important issues in 
astrophysics \cite{bethe,lat01,bom,eng,pra}. For example, 
the nuclear symmetry energy plays an important role in understanding the 
nucleosynthesis during the pre-supernova evolution of massive stars, 
the mechanisms for supernova explosions, the cooling rate of protoneutron 
stars and the associated neutrino fluxes, and the kaon condensation as well 
as the hadron to quark-gluon plasma phase transitions in neutron stars 
\cite{bethe,lat01}. However, our knowledge of nuclear symmetry energy remains 
limited in spite of extensive theoretical studies \cite{bom,kut,kub02}. 
On the other hand, radioactive beams, particularly the very energetic ones 
to be available at the planned Rare Isotope Accelerator (RIA) and the new 
accelerator facility at the German Heavy Ion Accelerator Center (GSI), 
provide a great opportunity to study the density dependence of the nuclear 
symmetry energy. Significant progress has already been achieved in 
identifying the observables that are sensitive to the symmetry energy.
It was proposed that information about the nuclear symmetry energy can be 
extracted from measurements of the neutron-skins of radioactive nuclei 
via their total reaction cross sections \cite{oya} and those of stable 
heavy nuclei via parity-violating electron scatterings \cite{hor01}.  
Promising probes to the nuclear symmetry energy have also been found 
in heavy-ion collisions induced by neutron-rich nuclei, and 
these include the pre-equilibrium neutron/proton ratio \cite{li97}, the 
isospin fractionation \cite{fra1,fra2,xu00,tan01,bra02}, the isoscaling 
in multifragmentation \cite{betty}, the proton differential elliptic flow 
\cite{lis} and the neutron-proton differential transverse flow \cite{li00} 
as well as the $\pi ^{-}$ to $\pi ^{+}$ ratio \cite{li02}. 

Since two-particle correlation functions, through final state interactions 
and quantum statistical effects, have been shown to be a sensitive probe 
to the space-time distributions of emitted particles in heavy-ion 
collisions \cite{bauer}, it is of interest to investigate if they can 
also be used to study the density dependence of the nuclear symmetry energy. 
In this Letter, we show that the emission times of neutrons and protons 
are indeed sensitive to the nuclear symmetry energy. A stronger density 
dependence in the nuclear symmetry energy leads to an earlier and more 
correlated emission of pre-equilibrium neutrons and protons. Consequently, 
strengths of the correlation functions for nucleon pairs with high total 
momenta, especially for neutron-proton pairs with low relative momenta,
are larger for a stiffer symmetry energy. Measurements of two-nucleon 
correlation functions in heavy ion collisions thus provides another possible 
tool for extracting useful information about the nuclear symmetry energy. 

Our study is based on an isospin-dependent Boltzmann-Uehling-Uhlenbeck 
(IBUU) transport model (e.g., Refs. \cite{buu,li97,li02,li00,li96}). 
In this model, the initial positions of protons and neutrons in the 
colliding nuclei are determined according to their density distributions 
predicted by the relativistic mean-field (RMF) theory. Their initial momenta 
are then taken to have a uniform distribution inside the neutron or proton 
Fermi sphere with its Fermi momentum determined from the local density using 
the Thomas-Fermi approximation. For the isoscalar potential, we use as 
default the Skyrme potential with an incompressibility $K_{0}=380$ MeV. 
This potential has been shown to approximately reproduce the transverse 
flow data in heavy-ion collisions, although the latter are best described 
by a momentum-dependent soft potential with $K_{0}=210$ MeV 
\cite{pan93,zhang93}. The isospin effects are included through the
isospin-dependent total and differential nucleon-nucleon cross sections, 
the different Pauli blockings for protons and neutrons, the symmetry 
potential, and the Coulomb potential for protons.  For nucleon-nucleon 
cross sections, we use as default the experimental values in free space, 
in which the neutron-proton cross section is about a factor of 3 larger 
than the neutron-neutron or proton-proton cross section. For a review of 
the IBUU model, we refer the reader to Ref. \cite{li98}. 

\begin{figure}[ht]
\includegraphics[scale=1.1]{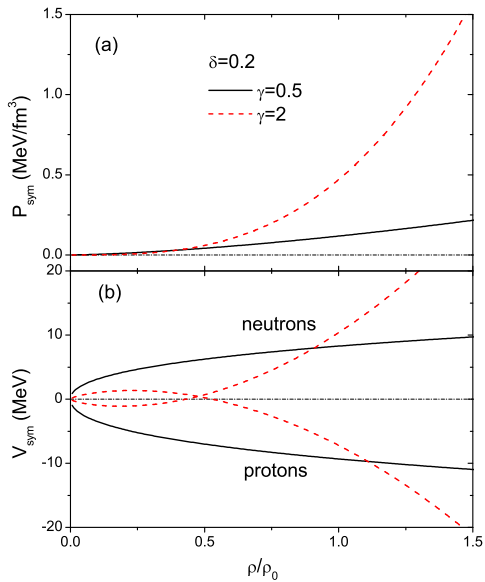}
\vspace{-1cm}
\caption{{\protect\small The symmetry pressure (upper panel) and 
potential (lower panel) in nuclear matter with an isospin asymmetry 
$\delta=0.2$ for the soft (solid curves) and stiff (dashed curves) 
symmetry energies.}}
\label{sympot}
\end{figure}

The energy per nucleon in an asymmetric nuclear matter is usually expressed as
\begin{equation}
E(\rho ,\delta )=E(\rho ,\delta =0)+E_{\text{\textrm{sym}}}(\rho )\delta
^{2}+{\cal O}(\delta^4),  
\label{eq1}
\end{equation}%
where $\rho =\rho _{n}+\rho _{p}$ is the baryon density; $\delta =(\rho
_{n}-\rho _{p})/(\rho _{p}+\rho _{n})$ is the relative neutron excess; and $%
E(\rho ,\delta =0)$ is the energy per particle in a symmetric nuclear matter,
while $E_{\rm sym}(\rho)$ is the nuclear symmetry energy. For the latter,   
we adopt the parameterization used in Ref. \cite{hei00} for studying 
the properties of neutron stars, i.e.,  
\begin{equation}
E_{\rm sym}(\rho )=E_{\rm sym}(\rho_0)\cdot u^{\gamma },
\end{equation}  
where $u\equiv \rho /\rho _{0}$ is the reduced density and 
$E_{\rm sym}(\rho_0)=35$ MeV is the bulk symmetry energy at normal nuclear 
matter density $\rho _{0}=0.16$ fm$^{-3}$. In the following, we consider 
the two cases of $\gamma=0.5$ (soft) and 2 (stiff) to explore the large 
range of $E_{\rm sym}(\rho)$ predicted by different theoretical models
\cite{bom}. The pressure due to the symmetry energy, given by 
$P_{\rm sym}=(\rho^2\partial E_{\rm sym}/\partial\rho)\delta^2$, 
in asymmetric nuclear matter with an isospin asymmetry 
$\delta=0.2$ are shown in the upper 
panel of Fig. \ref{sympot} for the two symmetry energies.  It is seen 
that the stiff symmetry energy leads to a larger pressure in asymmetric 
nuclear matter than that from the soft symmetry energy at the same density. 

The symmetry potential acting on a nucleon derived from the above nuclear
symmetry energy is \cite{lis}
\begin{eqnarray}\label{vasy}
V_{\rm sym}&=&\pm 2[E_{\rm sym}(\rho_0)u^{\gamma}-12.7u^{2/3}]\delta
\nonumber\\
&&+[E_{\rm sym}(\rho_0)(\gamma-1)u^{\gamma}+4.2u^{\frac{2}{3}}]\delta^{2},
\end{eqnarray}
where ``+" and ``-" are for neutrons and protons, respectively. In the 
lower panel of Fig. \ref{sympot}, the symmetry potentials for 
protons and neutrons in an isospin asymmetric nuclear matter with 
$\delta=0.2$ are shown. The symmetry potential is seen to be attractive 
for protons and repulsive for neutrons and increases with density,
except for the stiff symmetry potential at densities below $0.5\rho_0$.

\begin{figure}[ht]
\includegraphics[scale=0.9]{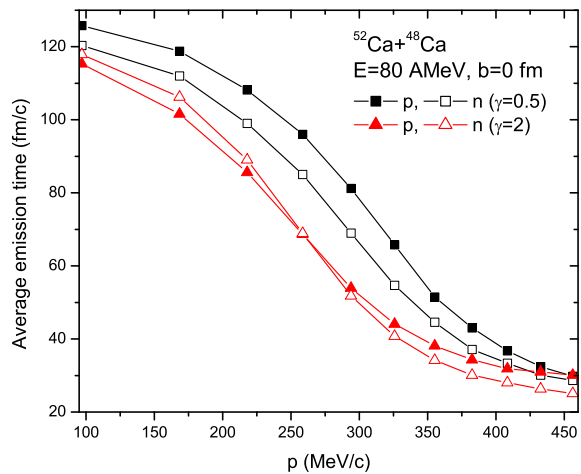}
\vspace{-1cm}
\caption{{\protect\small The average emission times of protons or neutrons 
as functions of their momenta for soft (solid curves) and stiff (dashed 
curves) symmetry energies.}}
\label{time}
\end{figure}

As an example, we study in this letter central collisions of $^{52}$Ca 
+ $^{48}$Ca at E=80 MeV/nucleon. This reaction system with an
isospin asymmetry $\delta=0.2$ can be studied at RIA. Nucleons are 
considered as being emitted when their local densities are less than 
$\rho _{0}/8$ and subsequent interactions with the mean field do not 
cause their recapture into regions of higher density. Other emission 
criteria, such as taking the nucleon emission time as its last collision 
time in the IBUU model, do not change our conclusions. Shown in Fig. 
\ref{time} are the average emission times of protons and neutrons as 
functions of their momenta. It is seen that the average emission time 
of nucleons with a given momentum is earlier for the stiff symmetry energy 
than for the soft one, as nucleon emissions are mainly governed by the 
pressure of the excited matter created during the collisions \cite{pawel,lis},
which is larger for the stiffer symmetry energy than for the soft one.
Moreover, there is a significant delay in the emission of protons 
when the symmetry energy is soft. This is due to the fact that the 
symmetry potential is generally repulsive for neutrons and attractive 
for protons, and their magnitudes at low densities, where most nucleons 
are emitted, are larger for the soft symmetry energy than for the stiff 
symmetry potential as seen in Fig. \ref{sympot}. The relative emission 
times of neutrons and protons in the case of the stiff symmetry energy 
depends on the density at which they are emitted, as the stiff symmetry 
potential changes sign when the nuclear density is below $0.5\rho_0$. 
It is interesting to note that high momentum nucleons are emitted during the
early pre-equilibrium stage of the collision when the density is relatively 
high, while low momentum ones are mainly emitted when the system is close 
to equilibrium and the density is low. 

It is well known that the space-time distributions of emitted particles 
can be extracted from the two-particle correlation functions; see, e.g., 
Refs. \cite{Boal90,bauer,Wied99}, for reviews. The two-proton correlation 
function has been studied most extensively. Recently, experimental data on 
two-neutron and neutron-proton correlation functions have also become 
available, and this has made it possible to deduce the relative emission 
times of neutrons and protons \cite{Ghetti}. Theoretically, the two-nucleon 
correlation function can be evaluated in the standard Koonin-Pratt formalism 
\cite{koonin77,pratt87} by convoluting the emission function 
$g(\mathbf{p},x)$, i.e., the probability for emitting a particle with 
momentum $\mathbf{p}$ from space-time point $x=(\mathbf{r},t)$, with the 
relative wave function of the two particles, i.e.,
\begin{equation}
C(\mathbf{P},\mathbf{q})=\frac{\int d^{4}x_{1}d^{4}x_{2}g(\mathbf{P}%
/2,x_{1})g(\mathbf{P}/2,x_{2})\left| \phi (\mathbf{q},\mathbf{r})\right| ^{2}
}{\int d^{4}x_{1}g(\mathbf{P}/2,x_{1})\int d^{4}x_{2}g(\mathbf{P}/2,x_{2})}.
\label{Eq1}
\end{equation}
In the above, $\mathbf{P(=\mathbf{p}_{1}+\mathbf{p}_{2})}$ and
$\mathbf{q(=}\frac{1}{2}(\mathbf{\mathbf{p}_{1}-\mathbf{p}_{2}))}$ are,
respectively, the total and relative momenta of the particle pair; and 
$\phi (\mathbf{q},\mathbf{r})$ is their relative wave function 
with $\mathbf{r}$ being their relative position, i.e., 
$\mathbf{r=(r}_{2}\mathbf{-r}_{1}\mathbf{)-}$ 
$\frac{1}{2}(\mathbf{\mathbf{v}_{1}+\mathbf{v}_{2})(}t_{2}-t_{1}\mathbf{)}$. 

\begin{figure}[ht]
\includegraphics[scale=1.1]{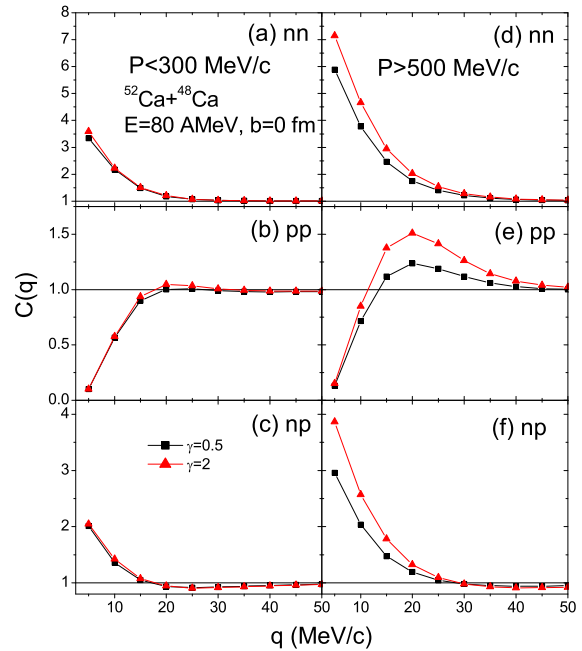}
\vspace{-1cm}
\caption{{\protect\small Two-nucleon correlation functions gated on the 
total momentum of nucleon pairs. Left panels are for $P<300$ MeV/c while 
right panels are for $P>500$ MeV/c.}}
\label{gatedCF}
\end{figure}

Using the program Correlation After Burner \cite{hbt}, which takes into 
account final-state nucleon-nucleon interactions, we have evaluated 
the two-nucleon correlations from the emission function given by the 
IBUU model. Shown in Fig. \ref{gatedCF} are the two-nucleon correlation 
functions gated on the total momentum ($P$) of nucleon 
pairs from central collisions of $^{52}$Ca + $^{48}$Ca at E=80 MeV/nucleon. 
The left and right panels are for $P<300$ MeV/c and 
$P>500$ MeV/c, respectively. For both neutron-neutron and 
neutron-proton correlation functions, they peak at $q\approx 0$ MeV/c. 
The proton-proton correlation function is, on the other hand, peaked at 
about $q=20$ MeV/c due to the strong final-state s-wave attraction 
but is suppressed at $q=0$ as a result of Coulomb repulsion and 
wave-function antisymmetrization between the two protons. 
Since the emission times of low momentum nucleons are not affected much 
by the different symmetry energies used in the IBUU model as shown 
in Fig. \ref{time}, the two-nucleon correlation functions are thus 
similar for the stiff and soft symmetry energies. On the other hand, 
the emission times of high momentum nucleons, which are dominated by 
those with momenta near 250 MeV/c, differ appreciably for the two 
symmetry energies considered here. The correlation functions of these nucleon 
pairs thus show a strong dependence on symmetry energy. Gating on
nucleon pairs with high total momentum thus allows one to select those 
nucleons that have stronger correlations as a result of small  
spatial separations at emissions. From Fig. \ref{gatedCF}, it is seen 
that the correlation functions for neutron-neutron and neutron-proton 
pairs with high total momentum but low relative momentum of $q=5$ MeV/c
are, respectively, about 20\% and 30\% higher for the stiff symmetry energy 
than for the soft symmetry energy, while the correlation function for 
proton-proton pairs with high total momentum and relative momentum of
$q=20$ MeV/c is about 20\% higher for the stiff symmetry energy than for
the soft symmetry energy. The neutron-proton correlation function thus 
has the largest sensitivity to nuclear symmetry energy. As shown in Fig. 
\ref{time} and discussed earlier, the relative emission times of neutrons 
and protons is sensitive to nuclear symmetry energy. Since the 
pre-equilibrium neutrons and protons are emitted almost simultaneously
in the case of the stiff symmetry energy, they are strongly 
correlated, leading thus to a larger value for the neutron-proton correlation 
function. On the other hand, protons and neutrons are less correlated
in the case of the soft symmetry energy as proton emissions are delayed 
relative to neutron emissions. As a result, the soft symmetry energy
gives a smaller value for the neutron-proton correlation function.  
Furthermore, with the stiff symmetry energy, neutrons as well as 
protons have smaller spatial separations at emissions as they are 
emitted earlier during the collision. Therefore, neutrons and also protons 
are more correlated among themselves for the stiff symmetry energy 
than for the soft one. Our results thus clearly demonstrate
that correlation functions of nucleon pairs with high total 
momentum can indeed reveal sensitively the effect of the nuclear
symmetry energy on the space-time distributions of nucleons at emissions.

We have also studied the dependence of two-nucleon correlation functions 
on the incompressibility $K_0$ and nucleon-nucleon cross sections.
We find that changing the value of $K_0$ from 380 to 210 MeV or using 
the in-medium nucleon-nucleon cross sections predicted by the 
Dirac-Brueckner approach based on the Bonn A potential \cite{lgq9394}, 
which has a smaller magnitude and weaker isospin dependence than 
the free ones but a strong density dependence, changes both the neutron-proton 
and proton-proton correlation functions by about 5\%, similar to those 
found in Ref. \cite{bauer}. The weak dependence of the two-nucleon 
correlation functions on the stiffness of isoscalar energy is due to the 
reduced difference in the pressure of the excited matter as a result of 
different maximum densities reached in the collision, with the stiff one 
giving a lower density than the soft one. Details on these results 
as well as the dependence of the symmetry energy effects studied here 
on the impact parameter, incident energy, and masses of the colliding system
will be reported elsewhere \cite{long}.

In conclusion, we have studied the two-nucleon correlation functions
in intermediate-energy heavy-ion collisions induced by neutron-rich
nuclei in the framework of an isospin-dependent transport model. 
We find that the nuclear symmetry energy affects significantly the 
emission times of neutrons and protons. A stiffer symmetry energy
leads to an earlier and nearly simultaneous emission of pre-equilibrium 
neutrons and protons. For the soft symmetry energy, nucleon emissions
are delayed compared to that for the stiff symmetry energy, and the
delay is longer for protons than for neutrons. As a result, the correlation 
functions for nucleon pairs with high total momenta, especially for 
neutron-proton pairs with low relative momenta, are stronger for a stiffer 
symmetry energy. Studies of two-nucleon correlation functions in heavy ion 
collisions thus provides a possible tool for extracting useful information 
about the density dependence of the nuclear symmetry energy.

This work is based on work supported by the U.S. National Science Foundation 
under Grant Nos. PHY-0088934 and PHY-0098805 as well as the Welch Foundation 
under Grant No. A-1358. The work of LWC is also supported by the National 
Natural Science Foundation of China under Grant No. 10105008, while that 
of VG is partially supported by a fellowship from the National Institute 
of Nuclear Physics (INFN) in Italy.

\end{document}